\RequirePackage{fix-cm}
\documentclass[smallextended]{svjour3}       % onecolumn (second format)
\smartqed  % flush right qed marks, e.g. at end of proof
\usepackage{graphicx}
\usepackage{epsfig} %% for loading postscript figures
\usepackage[caption=false]{subfig}
\usepackage{amsmath}
\usepackage{float}
\usepackage{placeins}
\begin{document}

\title{A pressure strain correlation model employing extended tensor bases %\thanks{Grants or other notes
%about the article that should go on the front page should be
%placed here. General acknowledgments should be placed at the end of the article.}
}

%\titlerunning{Short form of title}        % if too long for running head

\author{J. P. Panda         \and
        H. V. Warrior %etc.
}

%\authorrunning{Short form of author list} % if too long for running head

\institute{J. P. Panda \at
              Department of Ocean Engineering and Naval Architecture
				IIT Kharagpur, India \\
              \email{jppanda@iitkgp.ac.in}           %  \\
%             \emph{Present address:} of F. Author  %  if needed
           \and
           H. V. Warrior \at
             Department of Ocean Engineering and Naval Architecture
IIT Kharagpur, India
}

\date{Received: date / Accepted: date}
% The correct dates will be entered by the editor

\maketitle

\begin{abstract}
Accurate and robust models for the pressure strain correlation are an essential component for the success of Reynolds Stress Models in turbulent flow simulations. However replicating the non-local action of pressure using only local tensors places a large limitation on potential model performance. In this paper we outline an approach that extends the tensor basis used for pressure strain correlation modeling to formulate models with improved precision and robustness. This set of additional tensors is analyzed and justified based on physics based arguments and analysis of simulation data. Using these tensors models for the rapid and slow pressure strain correlation are developed. The resulting complete pressure strain correlation model is tested for a wide variety of turbulent flows, while being contrasted against the predictions of established models. It is shown that the new model provides significant improvement in prediction accuracy.   
% \PACS{PACS code1 \and PACS code2 \and more}
% \subclass{MSC code1 \and MSC code2 \and more}
\end{abstract}

\section{Introduction}
\label{intro}
Turbulent flows appear in problems of interest to many fields of engineering sciences e.g. aeronautics, mechanical, chemical engineering and in oceanographic, meteorological and astrophysical sciences, besides others. Improved predictions about the evolution of these flows would lead to significant advances in all these fields of science and engineering. 

In academic and industrial applications, most investigations into turbulent flow problems use turbulence models. Turbulence models are simplified relations that express quantities that are difficult to compute in terms of simpler flow parameters. They relate higher-order unknown correlations to lower-order quantities. These unknown correlations represent the actions of viscous dissipation, pressure-velocity interactions, etc. For example pressure strain correlation is a non-local phenomenon and is difficult to compute. Using models for pressure strain correlation, it is expressed as a function of Reynolds stresses, dissipation and mean velocity gradients which are local quantities. This enables us to estimate the pressure strain correlation and its effects on flow evolution in a simpler manner that is computationally inexpensive. Turbulence models are an essential component of all computational fluid dynamics software and are used in almost all simulations into real life fluid flows of engineering importance. 

A majority of industrial applications use simple two-equation turbulence models like the $k-\epsilon$ and $k-\omega$  models. However recent emphasis in the scientific community has shifted to Reynolds stress models (\cite{hanjalic2011,durbin2017,klifi2013,mishra3,jakirlic2015,manceau2015,eisfeld2016,schwarzkopf2016,moosaie2016,lee2016,mishra6,sun2017}). Reynolds stress models have the potential to provide better predictions than turbulent viscosity based models at a computational expense significantly lower than DNS studies. They may be able to model the directional effects of the Reynolds stresses and additional complex interactions in turbulent flows (\cite{johansson1994}). They have the ability to accurately model the return to isotropy of decaying turbulence and the behavior of turbulence in the rapid distortion limit (\cite{pope2000}).

Reynolds stress models use equations for the transport of the individual components of the Reynolds stress tensor. This Reynolds stress transport equation forms the foundation of the Reynolds stress modeling approach and is given by \cite{pope2000}
\begin{equation}
\begin{split}
&\partial_{t} \overline{u_iu_j}+U_k \frac{\partial \overline{u_iu_j}}{\partial x_k}=P_{ij}-\frac{\partial T_{ijk}}{\partial x_k}-\eta_{ij}+\psi_{ij},\\
&\mbox{where},\\ 
& P_{ij}=-\overline{u_ku_j}\frac{\partial U_i}{\partial x_k}-\overline{u_iu_k}\frac{\partial U_j}{\partial x_k},\\
&\ T_{kij}=\overline{u_iu_ju_k}-\nu \frac{\partial \overline{u_iu_j}}{\partial{x_k}}+\delta_{jk}\overline{ u_i \frac{p}{\rho}}+\delta_{ik}\overline{ u_j \frac{p}{\rho}},\\
&\eta_{ij}=-2\nu\overline{\frac{\partial u_i}{\partial x_k}\frac{\partial u_j}{\partial x_k}}  \\
&\psi_{ij}= \overline{\frac{p}{\rho}(\frac{\partial u_i}{\partial x_j}+\frac{\partial u_j}{\partial x_i})}\\
\end{split}
\end{equation}
The turbulence production process is represented by $P_{ij}$ and represents transfer of energy from the mean velocity field to the turbulent fluctuations. $\eta_{ij}$ represents the dissipation process where the turbulent kinetic energy is lost as internal energy. The turbulent transport process is represented by $T_{ijk}$ and has contributions from viscous diffusion, pressure transport and turbulent convection. Finally $\psi_{ij}$ represents the pressure strain correlation and redistributes turbulent kinetic energy among the components of the Reynolds stresses. Of these terms, production is the only process that is closed at the single point level. The other terms require models for their closure. The accuracy of the Reynolds stress modeling approach depends on the quality of the models developed for these turbulence processes. Out of these the modeling of the pressure strain correlation is often considered to be the most important.

The pressure strain correlation of turbulence consists of two components: the slow pressure strain correlation modeling the non-linear interactions in between the fluctuating velocity field and the rapid pressure strain correlation modeling the interactions between the mean velocity and the fluctuating velocity field. This can be seen in the Poisson equation for pressure \cite{pope2000}
\begin{equation}
\frac{1}{\rho}{\nabla}^2({p}^{R}+{p}^{S})=-2\frac{\partial{U}_j}{\partial{x}_i}\frac{\partial{u}_i}{\partial{x}_j}-\frac{\partial^2 u_iu_j}{\partial x_i \partial x_j}
\end{equation}
Here $p^R$ and $p^S$ are the rapid and slow components of pressure. On the right-hand side of Eqs. (2), the first term represents linear interactions between the fluctuating velocity field with the mean velocity gradient and the second term represents the non-linear interactions in between the fluctuating velocity field.

Due to its importance, there have been many attempts to develop closure models for the pressure strain correlation. \cite{chou} established the formulation for the second moment closure approach and introduced the pressure strain correlation term. \cite{rotta1951}  developed a linear closure for the slow pressure strain correlation term using a modeling expression that was linear in the Reynolds stresses. \cite{lrr} developed a model for the complete pressure strain correlation. They developed a novel closure for the rapid pressure term and incorporated the model of \cite{rotta1951} for the slow pressure strain correlation. \cite{jones1984} attempted to develop pressure strain correlation models that could be applicable for complex recirculating flows. Their model expression was similar to \cite{lrr} but the closure coefficients were calibrated to different values determined by the best agreement with their data for high Reynolds number homogeneous flows. \cite{ssmodel} developed a nonlinear extension for the slow pressure‐strain correlation for high Reynolds number flows. This model was able to show improved agreement with the non-linear trends in the return to isotropy behavior. This was extended to a fully non-linear quadratic model for the complete pressure strain correlation in \cite{ssg}. \cite{johansson1994} formulated a non-linear model for the rapid pressure strain correlation with quartic terms. This model showed improved agreement for some homogeneous turbulent flows.   

In spite of these modeling developments, there remain deficiencies in the performance of established models for the pressure strain correlation. These deficiencies are two-fold: limitations in accuracy and limitations in realizability.

Established pressure strain correlation models have unsatisfactory accuracy in some important classes of flows. For example in vorticity dominated flows their predictions may not be satisfactory. For these flows linear stability theory, experiments and DNS show growth in the turbulent kinetic energy. However established models predict that turbulence is decaying in these cases \cite{bns}. Similarly the predictions of available pressure strain correlation models are often inadequate in non-equilibrium turbulent flows, flows with swirl and re-circulation, etc \cite{mishra4}. 

Established pressure strain correlation models suffer from realizability issues. Realizability conditions ensure that the predictions of the turbulence model are consistent with a random stochastic process. The pressure strain correlation models available presently lead to realizability violations at or in the neighborhood of the two-component limit of turbulence. While the two-component limit is termed as a limiting state for the Reynolds stresses, it is found in many engineering flows. For example in near wall turbulence the state of the Reynolds stress tensor is extremely close to the two-component limit with the wall normal component of the Reynolds stresses being negligible. Such realizability violations in important flows limit the applicability of pressure strain correlation models.   

Most classical pressure strain correlation models have focused on the closure modeling expression and the values of the closure coefficients to improve the performance of models. With respect to the model expression there has been a trend toward more complex terms that are non-linear in the Reynolds stress tensor \cite{ssg}. For example while the model of \cite{lrr} was linear in the Reynolds stress tensor, the model of \cite{ssg} is quadratic and the model of \cite{johansson1994} is quartic. With respect to the closure coefficients, investigations have tried to calibrate them to more specialized data sets from experiments and DNS. Investigators have also made the closure coefficients functions of the invariants of the Reynolds stress tensor. This allows additional degrees of freedom in the modeling expression and enables better agreement with additional data sets. However the improvements due to such steps have been incremental. The central issues of unsatisfactory accuracy in specific important classes of flow or the issues with realizability are still present and important. 

Some investigations have raised questions about the inadequacy of the modeling basis used in pressure strain correlation closures. The modeling basis is composed of the set of tensors used in the modeled constitutive equation for the pressure strain correlation. In classical one-point closure modeling these are one-point tensors including the Reynolds stress anisotropies, the turbulent kinetic energy and the dissipation. The set of tensors used in the modeling basis determines the type and extent of information about the turbulent flow field that is available in the model formulation. In an incompressible flow pressure is governed by the Poisson equation. Due to the elliptic nature of this governing equation the pressure strain correlation is not a one-point tensor and attempts to model it using one-point tensors may be limited. \cite{kr1995} have claimed that in rotation-dominated turbulent flows, the modeling basis for the pressure strain correlation is limited. They introduced additional non-local tensors to the modeling basis like stropholysis, circulicity, etc. \cite{cambon1992} have also claimed that additional tensors may be needed to model the pressure strain correlation in rotation-dominated flows. However both these models use non-local tensors that may not be available in an engineering application. \cite{mishra1} and \cite{mishra2} have carried out a spectral analysis to outline the manner in which the modeling basis is limited and the manner in which it affects the ability of the model to replicate specific features of turbulent flows. 

If the limitations in the pressure strain correlation models are due to limited modeling basis, there are three important questions to be answered: 
\begin{enumerate}
\item What tensors need to be added to the modeling basis to have additional information that is relevant for modeling. 

\item While many different correlations and turbulent statistics may be added to the modeling basis and may offer different degrees of benefit, we must identify the optimal tensors to be added. 

\item Finally with these added tensors, how much improvement can we show in the performance of single-point pressure strain correlation models. 
\end{enumerate}

In this paper we address these questions in order. Using physics based arguments we outline a set of tensors to be added to the modeling basis for the slow pressure strain correlation and separately for the rapid pressure strain correlation. We show that these tensors add missing information to the modeling effort that is important to improve the potential performance of pressure strain correlation closures. We develop a complete model for the pressure strain correlation using this extended modeling basis. This model is tested for a range of mean flows while compared to DNS results. In this investigation, we use the popular models of \cite{lrr} and \cite{ssg} for comparison. The present model shows improved agreement with DNS results and significant improvements over these earlier pressure strain correlation models.

\section{Theoretical and mathematical details}
\label{sec:2}
In this section we outline our procedure to select specific tensors to the modeling basis for the pressure strain correlation. During this process, physical arguments for the choice of specific tensors and the particular benefits that they offer, with respect to the modeling of definite features of the pressure velocity interaction term. We demarcate this procedure sequentially, first for the slow pressure strain correlation model and then for the rapid pressure strain correlation model. During this selection, we try to consider tensors that are still single point and are available in the engineering single point modeling methodology. Following this selection, we develop the individual slow and rapid pressure strain correlation models with this expanded modeling basis. 

\subsection{Slow pressure strain correlation modeling basis}
Considering the slow pressure strain correlation model, we commence with the details of the rate of dissipation tensor. The rate of dissipation tensor can be decomposed into its deviatoric and isotropic components:
\begin{equation}
\eta_{ij}=D_{ij}+\frac{2}{3}\eta \delta_{ij}
\end{equation}

Here, $\eta=\frac{\eta_{ii}}{2}$ and $D_{ij}$ is the deviatoric component of the rate of dissipation tensor.

Traditionally, The deviatoric component of the rate of dissipation tensor is combined with the pressure strain correlation mechanism and the two are modeled together \cite{lumley1977}
\begin{equation}
\psi_{ij}=D_{ij}+\psi^{'}_{ij}
\end{equation}

In flows where the Reynolds number is large dissipation can be assumed to be isotropic $D_{ij}=0$. In most Reynolds Stress Modeling investigations this assumption is adopted and it is assumed that the rate of dissipation tensor is nearly isotropic. For all practical modeling purposes, $\psi_{ij}$ is the slow pressure strain correlation only. However recent direct numerical simulation studies suggest that this  assumption is inadequate \cite{14,15}. For example in near wall turbulence this assumption is unsatisfactory \cite{15}. In fact \cite{16} have proved that if the large scale structures in a turbulent flow are anisotropic the small scale turbulent motions will have a significant level of anisotropy. Due to these arguments the assumption of the isotropy of the rate of dissipation is a significant shortcoming and causes deficiencies in the slow pressure strain correlation model. To address the shortcomings due to this assumption we introduce the dissipation anisotropy tensor ($d_{ij}$) in the modeling basis:
\begin{equation}
d_{ij}=\frac{\eta_{ij}}{\eta}-\frac{2}{3} \delta_{ij}
\end{equation}

This tensor allows the model to have information about the anisotropy in the rate of dissipation mechanism and should improve the predictions of the models especially in the inhomogeneous flows.

A considerable amount of information required for the closure modeling of the terms in the Reynolds Stress Models is contained in the two-point correlation tensor, $R_{ij}(\vec{x},\vec{r})=\left\langle u_i(\vec{x})u_j(\vec{x}+\vec{r})\right\rangle$. The two-point correlation contains significant information about the dissipation and the pressure strain correlation, both of which can be expressed as functionals of the two-point correlation. The two-point correlation also has important information about the turbulent length scales. As the two-point correlation is non-local it is not used in the single-point modeling basis. This causes another significant shortcoming in classical Reynolds Stress Models that is the assumption of a single integral length scale. This is markedly true in flows where the geometry of the flow domain or body forces lead to a co-ordinate direction in the flow being decidedly preferred. For example axisymmetric expansion and axisymmetric contraction mean flows. In many anisotropic turbulent flows, the characteristic length scale is observed to be varying in different directions \cite{panda2017,prandtl}. At the most basic level, we must try to include this anisotropy in the length scale in the modeling basis for the pressure strain correlation. We introduce the length scale anisotropy tensor ($l_{ij}$) in the modeling basis and derive it as follows. The length scale information tensor ($L_{ij}$) is defined as:
\begin{equation}
L_{ij}=\frac{3}{4}\frac{k^{3/2}}{\eta}(c^*_1 b_{ij} +c^*_2 d_{ij})
\end{equation}

A scaling factor $l$ is defined as $\frac{k^{3/2}}{\eta}$. The expression for the length scale anisotropy is given by
\begin{equation}
l_{ij}=\frac{L_{ij}}{l}=\frac{3}{4}(c^*_1 b_{ij} +c^*_2 d_{ij})
\end{equation}
This derivation can be found in detail in \cite{panda2017,prandtl}. Based on the investigation of \cite{panda2017}, the values of $c^*_1$ and $c^*_2$ are uniformly set to $\frac{4}{6}$. 

\subsection{Rapid pressure strain correlation modeling basis}

Considering the rapid (or linear) pressure strain correlation term, we address the level of information used to characterize the state of the turbulent flow field. One of the key shortcomings in the Reynolds Stress Modeling approach to pressure strain correlation closures is the use of only the Reynolds stress tensor to describe the state of the turbulent flow field. This leads to a coarse grained description that limits the potential accuracy of the rapid pressure strain correlation model. \cite{mishrathesis,mishra1,mishra2,mishra3,mishra4} have made important insights about the specific limitations due to this level of characterization of the turbulent flow field. They have shown that turbulent flow fields with the same Reynolds stresses can have very different internal structuring and lead to very different evolution \cite{mishra1,mishra2}. Using  spectral analysis, they have established a universal evolution (termed the statistically most likely behavior) that is dependent on the mean velocity gradient. This behavior was shown to be highly dependent on the mean velocity gradient of the flow \cite{mishra2,mishra3}. At the primary level, including information about the local mean velocity gradients may be a good substitute for detailed multi-point information about the internal structuring of the turbulent flow field. For information about the mean velocity gradient, we introduce the invariants of the mean velocity gradient in the modeling basis. In this paper we restrict ourselves to planar mean velocity gradients. Here the information about the mean velocity field can be included using the ellipticity parameter \cite{mishrathesis}:
\begin{equation}
\beta = \frac{{W}_{mn}{W}_{mn}}{{W}_{mn}{W}_{mn}+{S}_{mn}{S}_{mn}}
\end{equation}

\subsection{Integration of additional tensors into model expressions}

To integrate these three tensors into the model expression we adopt a practical recourse. For the slow pressure strain correlation, the addition of the tensors requires that the model expression be extended. On experimenting with variants where the coefficients of the closure expression were made functions of $d_{ij}$ and $l_{ij}$, we found the final model to not perform well. However the general expression for the rapid pressure strain correlation closure is retained and the closure coefficients are made functions of these three tensors. On experimenting with variations (where additional terms involving these tensors were included in the model expression) we have found that this does not negatively affect the performance of the new model. Additionally we hope that retaining the established closure expression and only changing the nature of the closure coefficients will encourage the scientific community to incorporate this model into their proprietary codes.  

A most general form of the slow pressure strain correlation can be written as:
\begin{equation}
\psi^{(s)}_{ij}=\beta_1 b_{ij} + \beta_2 (b_{ik}b_{kj}- \frac{1}{3}II_b \delta_{ij})
\end{equation}
Here $b_{ij}$ is the Reynolds stress anisotropy, $II_b$ is the second invariant of the Reynolds stress anisotropy tensor. In the most general case, $\beta_1$ and $\beta_2$ are assumed to be functions of the second and third invariants of the Reynolds stress anisotropy tensor. The model of \cite{rotta1951} assumed $\beta_1$ to be a constant and $\beta_2$ to be zero. \cite{ssmodel} assumed both $\beta_1$ and $\beta_2$ to be non-zero constants. The slow pressure-strain correlation used in this investigation has the model expression derived by \cite{panda2017}. This involves the most general expression for the slow pressure strain correlation in terms of the three tensors $b_{ij}, d_{ij}$ and $l_{ij}$.
\begin{equation}
\begin{split}
& \psi^{(s)}_{ij}=c_1 b_{ij} + c_2 d_{ij} + c_3 l_{ij} +c_4 (b_{ik}b_{kj}-\frac{1}{3}b_{mn}b_{mn} \delta_{ij}) \\ & +c_5 (b_{ik}d_{kj}- \frac{1}{3}b_{mn}d_{mn} \delta_{ij}) +c_6 (d_{ik}d_{kj}- \frac{1}{3}d_{mn}d_{mn} \delta_{ij}) \\
& +c_7 (b_{ik}l_{kj}- \frac{1}{3}b_{mn}l_{mn} \delta_{ij}) +c_8 (d_{ik}l_{kj}- \frac{1}{3}d_{mn}l_{mn} \delta_{ij}) \\
& +c_9 (l_{ik}l_{kj}- \frac{1}{3}l_{mn}l_{mn} \delta_{ij}).
\end{split}
\end{equation}
The values of the model coefficients have been derived in \cite{panda2017} and are given by $(c_1,c_2,c_3,c_4,c_5,c_6,c_7,c_8,c_9)=(3.1,1.1,-0.6,-4.3, -15.8, -7.2, 8.4, 6.6,9.8 )$. 

Considering the rapid pressure strain correlation, the linear form of the model expression is
\begin{equation}
\begin{split}
&\frac{\psi^{R}_{ij}}{k}=C_2 S_{ij} +C_3 (b_{ik}S_{jk}+b_{jk}S_{ik}-\frac{2}{3}b_{mn}S_{mn}\delta_{ij})+\\  
& C_4 (b_{ik}W_{jk} + b_{jk}W_{ik})
\end{split}
\end{equation}
Here $b_{ij}=\frac{\overline{u_iu_j}}{2k}-\frac{\delta_{ij}}{3}$ is the Reynolds stress anisotropy tensor, $S_{ij}$ is the rate of strain term for the mean velocity field and $W_{ij}$ is the rate of rotation term for the mean velocity field. Following the notation of \cite{ssg} $C_2, C_3$ and $C_4$ are the coefficients of the rapid pressure strain correlation model. 

\begin{figure}
\centering
\subfloat[$L_2$]{\includegraphics[height=5cm]{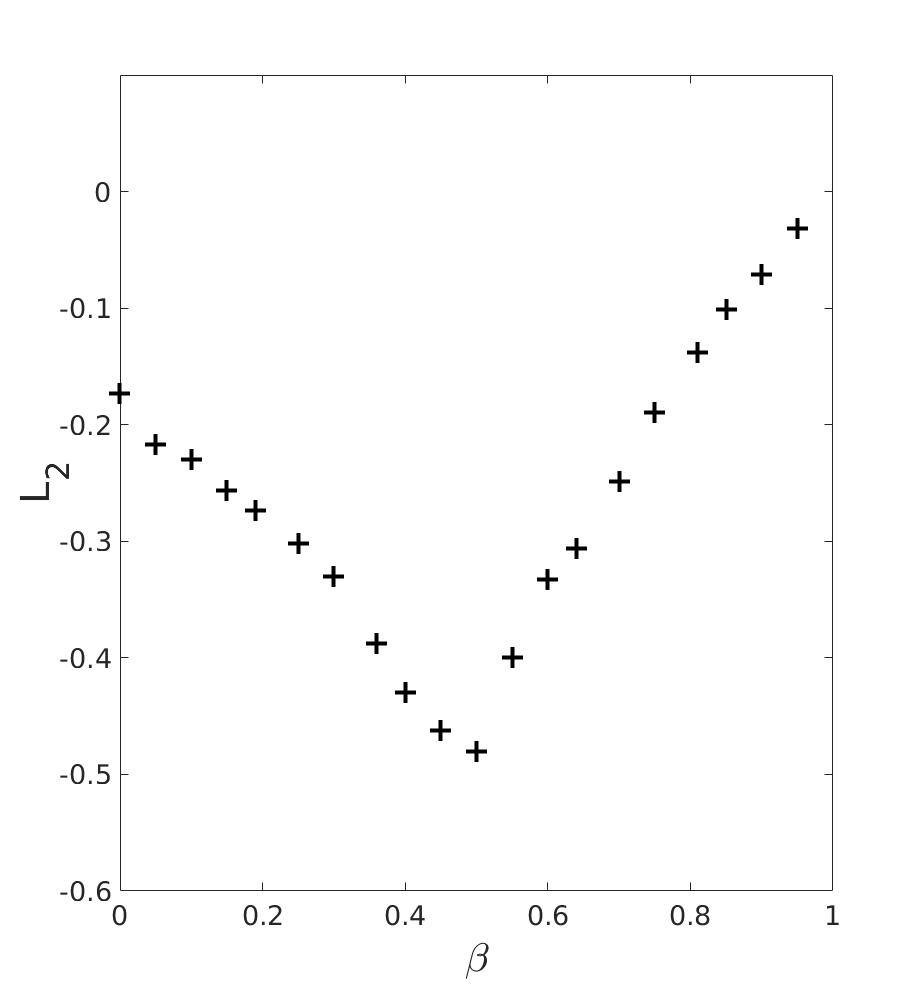}}
\subfloat[$L_3$]{\includegraphics[height=5cm]{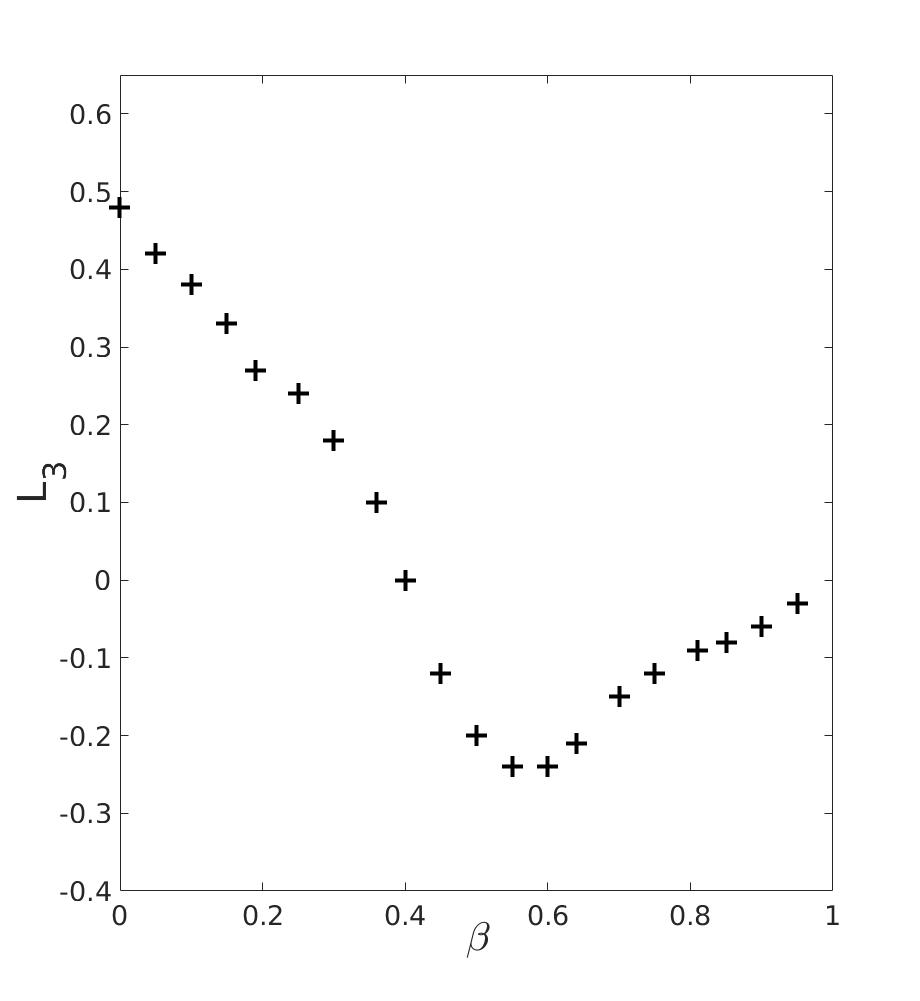}}
\subfloat[$L_4$]{\includegraphics[height=5cm]{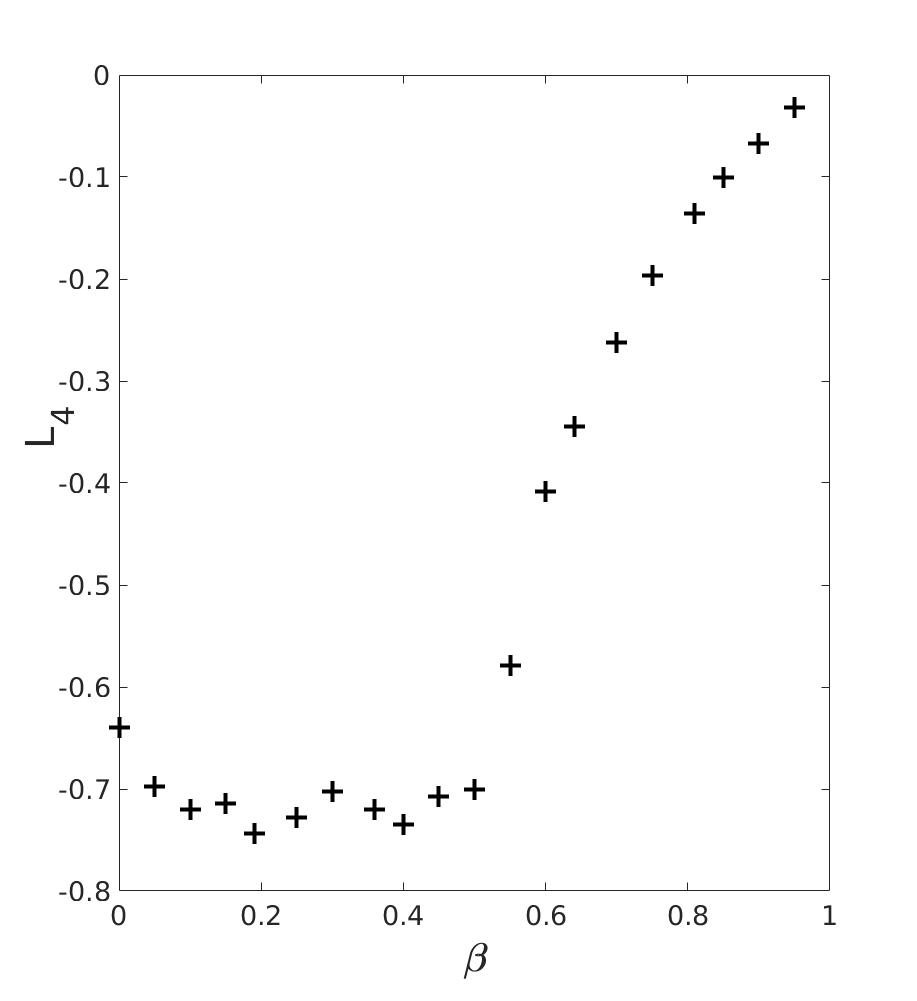}}
\caption{Calculated values of the rapid pressure strain correlation model coefficients $L_2, L_3$ and $L_4$ as functions of $\beta$.}
\end{figure}

Based on this form of the rapid pressure strain correlation, the Reynolds stress anisotropy evolution equation is derived from the Reynolds stress transport equation at the rapid distortion limit
\begin{equation}
\begin{split}
&\frac{d b_{ij}}{dt}= 2 b_{ij} b_{mn}S_{mn} + L_2 S_{ij}+\\
& L_3(b_{ik}S_{jk}+b_{jk}S_{ik}-\frac{2}{3}b_{mn}S_{mn}\delta_{ij})+L_4(b_{ik}W_{jk} + b_{jk}W_{ik})
\end{split}
\end{equation}
Here $L_2=\frac{C_2}{2}-\frac{2}{3}$, $L_3=\frac{C_3}{2}-1$ and $L_4=\frac{C_4}{2}-1$. Once the form of the rapid pressure strain correlation model expression is fixed, the modeling reduces to determine the values of the model coefficients $C_2, C_3$ and $C_4$ (or of $L_2, L_3$ and $L_4$). To integrate the additional tensor in the rapid pressure expression (the ellipticity parameter, carrying information about the mean velocity gradient) we follow the outline of \cite{mishra2,mishra3}. Here the model coefficients are made explicit functions of the ellipticity parameter. To this end, we use representation theory and try to ensure that the stationary state of the anisotropy evolution equation (Eq. (12)) matches the stationary state of the Reynolds stress anisotropy tensor observed in RDT simulations \cite{devaney2008}. Using representation theory the values of the Reynolds stress anisotropy at equilibrium can be expressed as a polynomial function in terms of the mean rate of strain and mean rate of rotation. Based on Pope \cite{pope1975}, the general form of this is given by 
\begin{equation}
b_{ij}=G_1 S_{ij}+ G_2(S_{ik}W_{kj}+W_{ki}S_{kj})+G_3(S_{ik}S_{kj}+\frac{(\beta-1)\delta_{ij}}{3})
\end{equation}
$G_1, G_2$ and $G_3$ are scalars that are functions of the invariants of flow statistics. This approach can be extended to three dimensional mean flow cases (\cite{mishra4}). In this paper, we study two dimensional mean flow cases that can be completely described using $\beta$. Using the polynomial form from Eqs. (13), and using a Matlab script to calculate values of the Reynolds stress anisotropies at the stationary equilibrium points (designated by $b_{11}^*$, $b_{22}^*$ and $b_{12}^*$), $G_1, G_2$ and $G_3$ can be expressed in terms of these stationary values of the Reynolds stress anisotropies 
\begin{equation}
\begin{split}
&G_1=\frac{b_{11}^*-b_{22}^*}{\sqrt{2(1-\beta)}} \\
&G_2=-\frac{b_{12}^*}{\sqrt{\beta(1-\beta)}}\\
&G_3=\frac{b_{11}^*+b_{22}^*}{(1-\beta)/3}
\end{split}
\end{equation}

At the stationary states for the Reynolds stress anisotropy, the evolution equation Eqs. (12) simplifies to 
\begin{equation}
\begin{split}
& 2 b_{ij}^* b_{mn}^*S_{mn} + L_2 S_{ij}+L_3(b_{ik}^*S_{jk}+b_{jk}^*S_{ik}-\frac{2}{3}b_{mn}^*S_{mn}\delta_{ij})\\
& +L_4(b_{ik}^*W_{jk} + b_{jk}^*W_{ik})=0
\end{split}
\end{equation}
Here $b_{ij}^*$ is the value of the Reynolds stress anisotropy at the stationary state. 

Replacing $b_{ij}^*$ in Eqs. (15) by the polynomial form from Eqs. (13), we get a equation for the coefficients $L_2, L_3, L_4$ as functions of $G_1, G_2, G_3$ 
\begin{equation}
\begin{split}
&L_2=-2(1-\beta)G_1^2-4\beta(1-\beta)G_2^2+\frac{(1-\beta)^2}{3}G_3^2 \\
&L_3=-(1-\beta)G_3\\
&L_4=2(1-\beta)G_2
\end{split}
\end{equation}

Finally substituting the values of $G_1, G_2$ and $G_3$ computed in Eqs. (14) into the Eqs. (16), we get the values of $L_2, L_3$ and $L_4$. So the steps of the formulation are as follows: 
\begin{enumerate}
\item Using Direct Numerical Simulations at the rapid distortion limit, we find the stationary values of $b_{ij}$ at a range of values of $\beta$.
\item Using these values, $b_{ij}^*$ in Eqs. (14), we find the values of $G_1;G_2;G_3$ as functions of $\beta$.  
\item Replacing these values of $G_1;G_2;G_3$ in Eqs. (16), at the Rapid Distortion limit, we can find the $L_2;L_3,L_4$ as functions of $\beta$. These are shown in Figure 1.
\item We use these values of $L_2, L_3, L_4$ into the calculation of $b_{ij}$ at lower strain rates as functions of $\beta$ using Eqs. (12), coupled with the model for the slow pressure strain correlation given in Eqs. (10).   
\end{enumerate}

At this point, we have outlined the additional tensors to be added to the modeling basis for the pressure strain correlation and the specific reasons for their addition. The slow and rapid pressure strain correlation models with these additional tensors have been formulated. In the next section we use these two model expressions together to simulate the evolution of general turbulent flows that are far off the limiting states of turbulence. This methodology follows the procedure counseled by \cite{speziale1992testing}, where they have warned against testing models of the rapid and slow pressure strain correlation in isolation. 

During the test cases, the turbulent kinetic energy ($k=\frac{\overline{u_lu_l}}{2}$) evolves as
\begin{equation}
\frac{dk}{dt}=P-\eta
\end{equation}
The modeled evolution equation for the dissipation is 
\begin{equation}
\frac{d\eta}{dt}=C_{\eta_1}\frac{\eta}{k}P - C_{\eta_2}\frac{\eta^2}{k}
\end{equation}
Here the values of the coefficients are taken as $C_{\eta_1}= 1.44$ and $C_{\eta_2}= 1.88$.

\section{Results and discussion}

\begin{figure}
\centering
\subfloat[$b_{11}$]{\includegraphics[height=5cm]{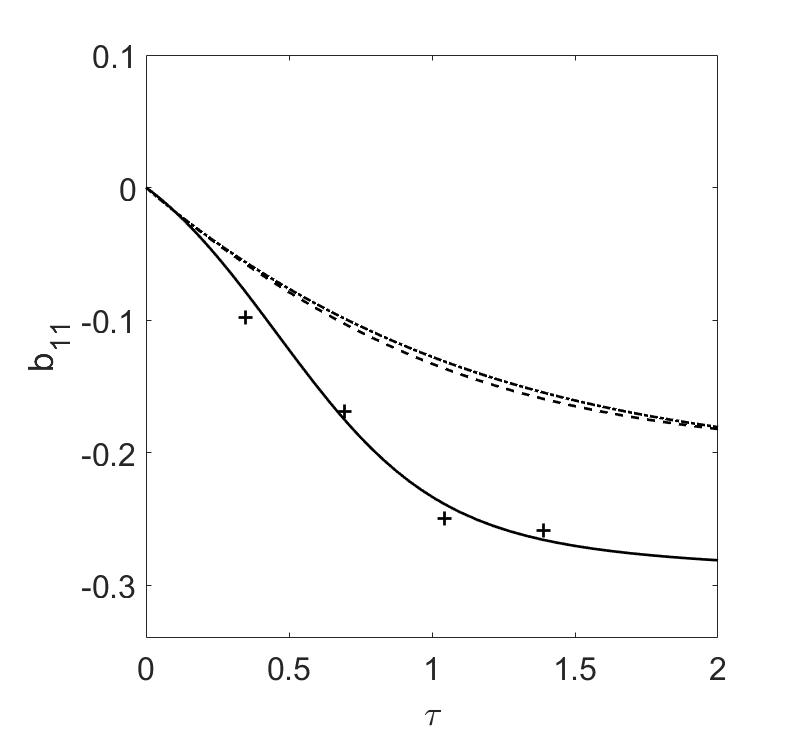}}
\subfloat[$k$]{\includegraphics[height=5cm]{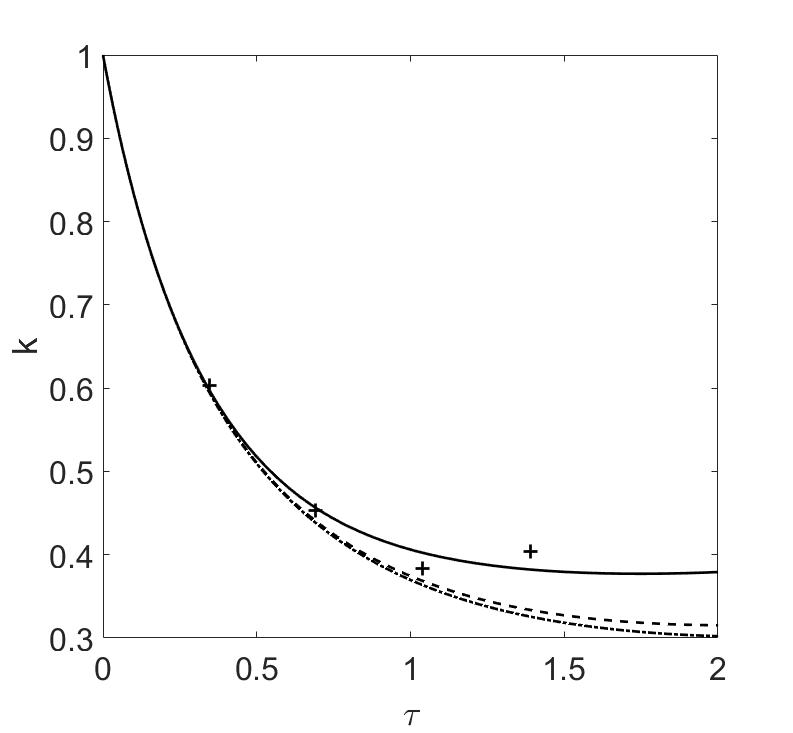}}
\caption{Evolution of a) the Reynolds stress anisotropy $b_{11}$ b) turbulent kinetic energy $k$ for plane strain mean flow. The predictions of the present model are shown by the solid line. SSG and LRR model are shown by the dashed and dash-dot lines. The data from the direct numerical
simulation of Lee and Reynolds \cite{lee1985} is included for comparison.}
\end{figure}

In this section, the present pressure strain correlation model is tested for a wide variety of general turbulent flows. We ensure that these flows are general in the sense that they are not at the limiting states of decaying turbulence or the rapid distortion limit. We use the predictions of established models by \cite{lrr} and \cite{ssg} as yard sticks to compare the performance of the present model. 

\begin{figure}
\centering
\subfloat[$k, E=1.5$]{\includegraphics[height=5cm]{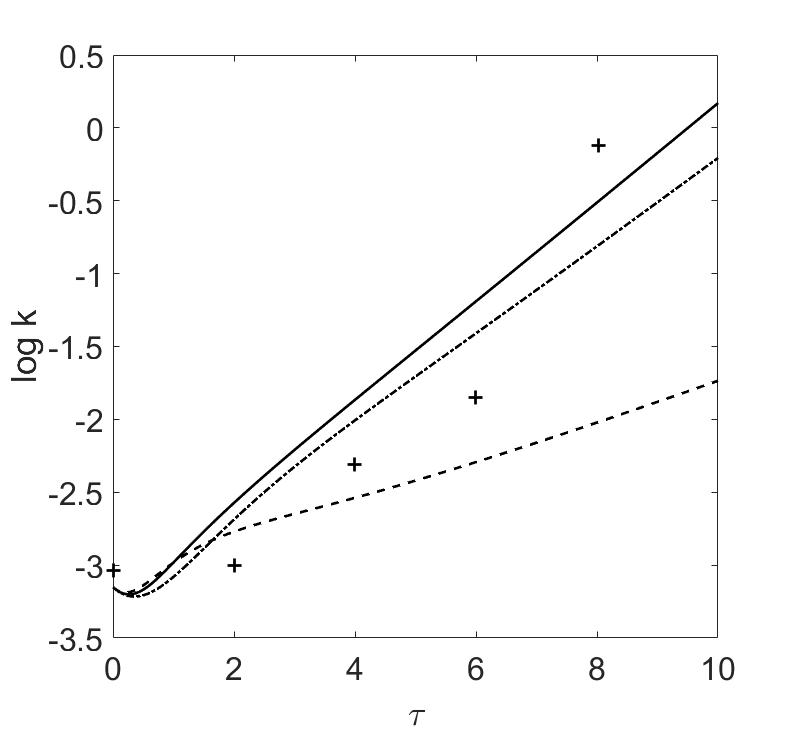}}\\
\subfloat[$k, E=2$]{\includegraphics[height=5cm]{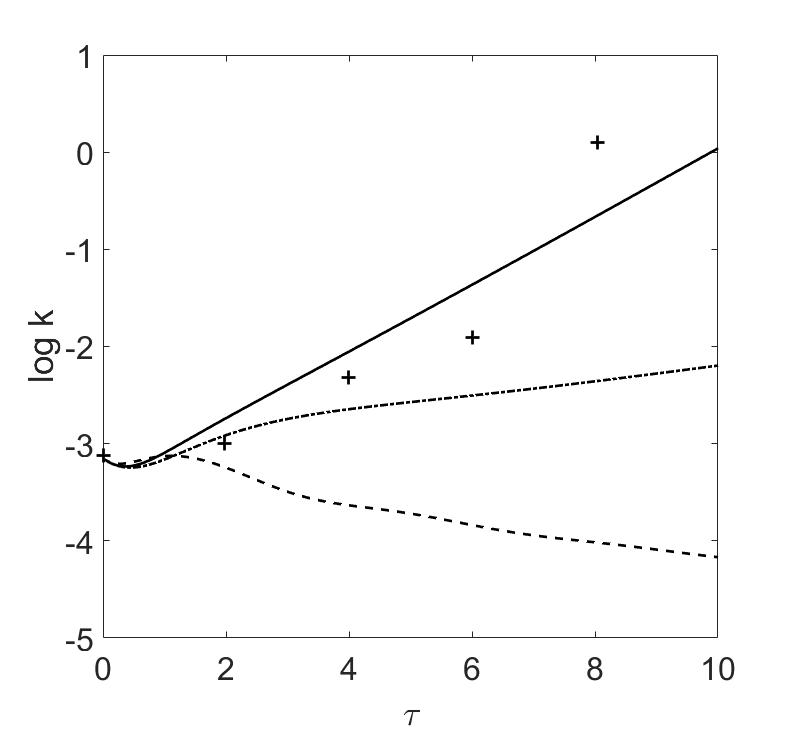}}\\
\subfloat[$k, E=3$]{\includegraphics[height=5cm]{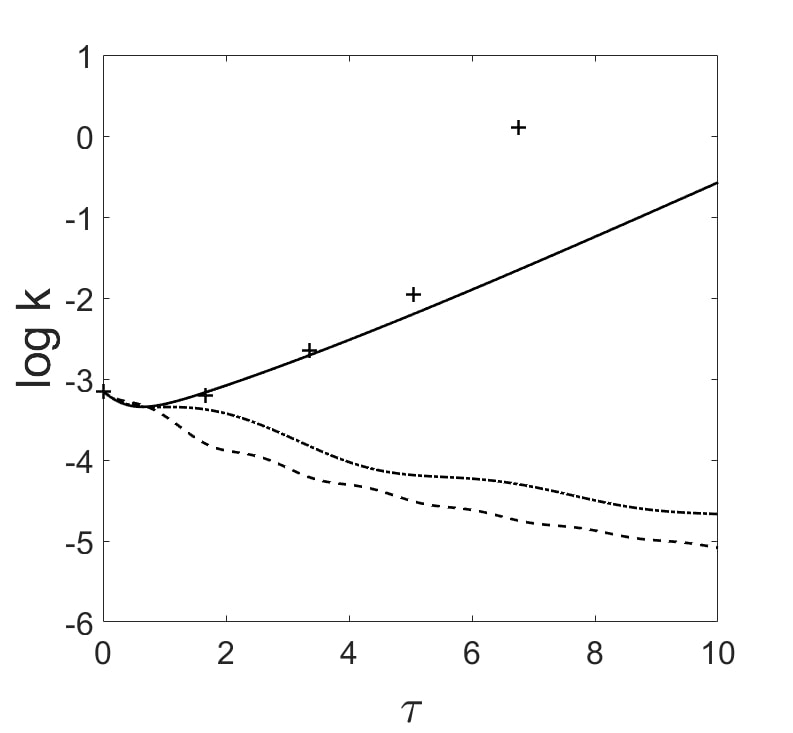}}\\
\caption{Turbulent kinetic energy evolution for elliptic flows a) E=1.5 b) E=2 c) E=3. The present model predictions are in the solid line,the SSG and the LRR model are shown in dash-dot and dotted lines. The data from the direct numerical simulation of  Blaisdell and Shariff \cite{bns} is included for comparison.}
\end{figure}

\begin{figure}
\centering
\subfloat[$b_{13}, E=1.5$]{\includegraphics[height=5cm]{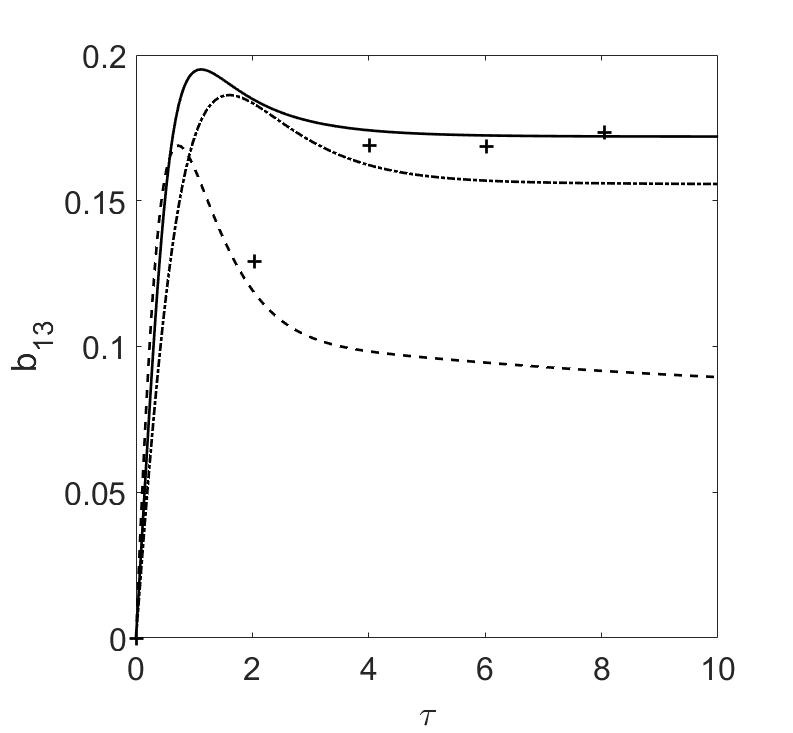}}\\
\subfloat[$b_{13}, E=2$]{\includegraphics[height=5cm]{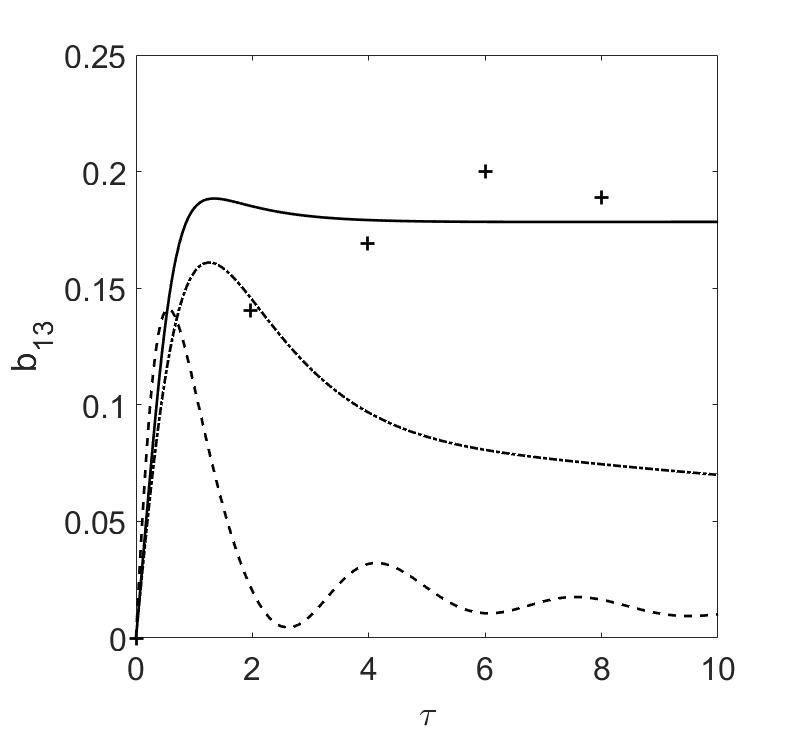}}\\
\subfloat[$b_{13}, E=3$]{\includegraphics[height=5cm]{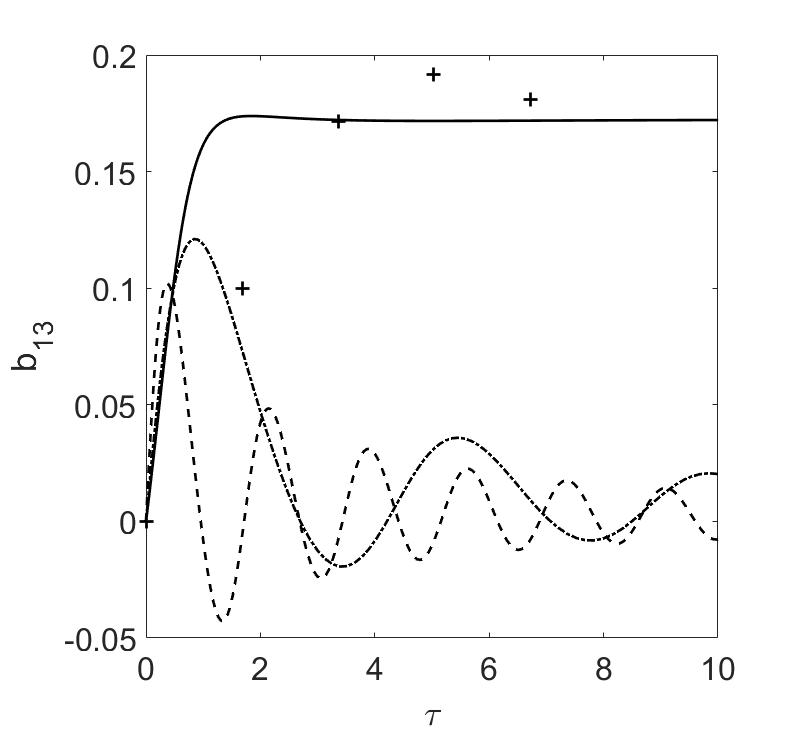}}\\
\caption{Reynolds stress anisotropy $b_{13}$ evolution for elliptic flows a) E=1.5 b) E=2 c) E=3. The present model predictions are in the solid line,the SSG and the LRR model are shown in dash-dot and dotted lines. The data from the direct numerical simulation of  Blaisdell and Shariff \cite{bns} is included for comparison.}
\end{figure}

In Figure 2 the evolution of Reynolds stress anisotropy and turbulent kinetic energy is shown for plane strain mean flow. The present model predictions are shown in a solid line, the SSG and the LRR model are shown in dash-dot and dotted lines respectively. DNS data from \cite{lee1985} is shown using unfilled circles in the figure. The predictions of the present model for both the components of the Reynolds stress anisotropy and the evolution of the turbulent kinetic energy show agreement with the DNS data. The present model is able to show some improvement in comparison to the predictions of popular models like those by \cite{lrr} and \cite{ssg}.

\begin{figure}
\centering
\subfloat[$k, S^*=3$]{\includegraphics[height=5cm]{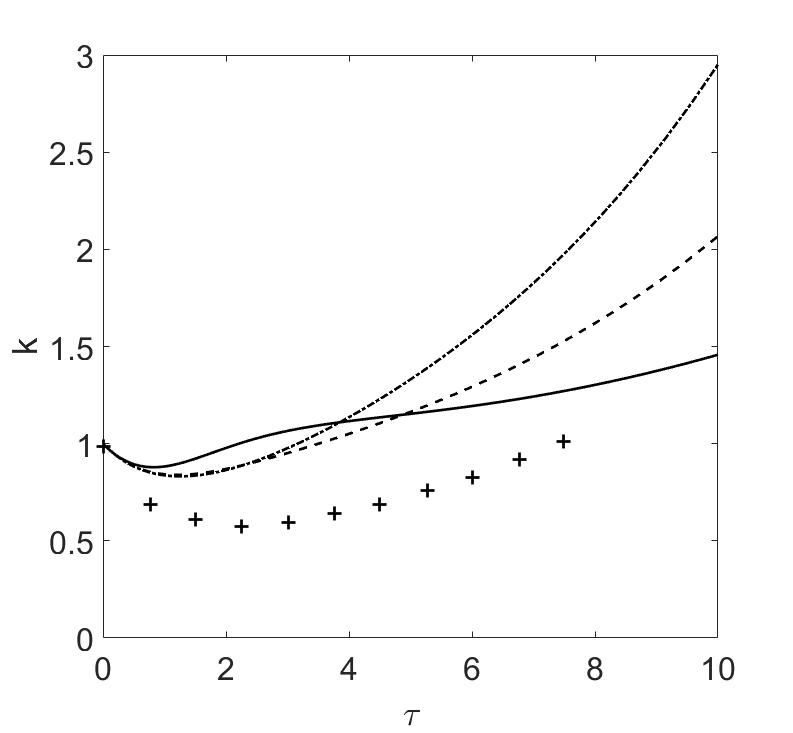}}\\
\subfloat[$k, S^*=15$]{\includegraphics[height=5cm]{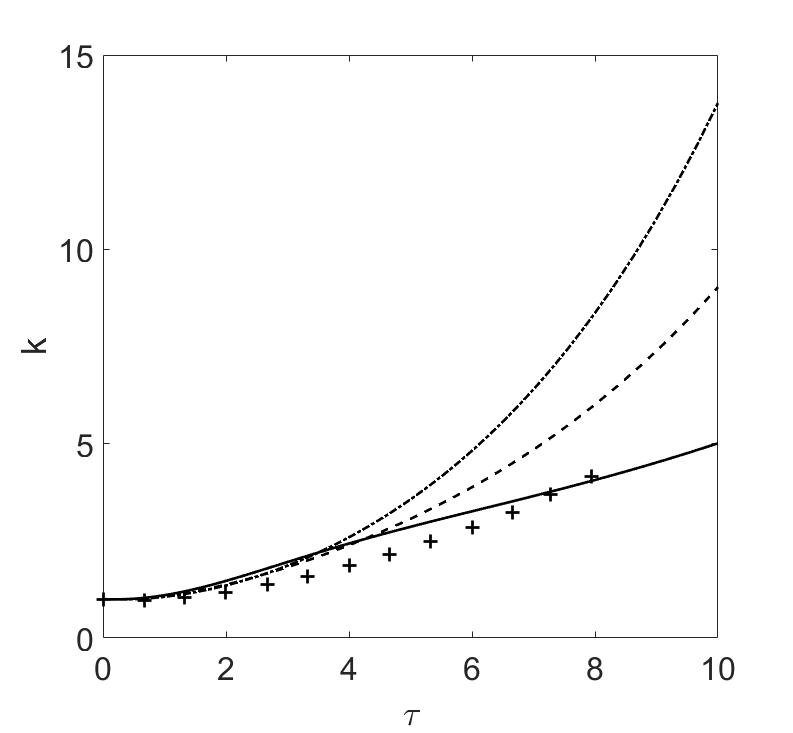}}\\
\subfloat[$k, S^*=27$]{\includegraphics[height=5cm]{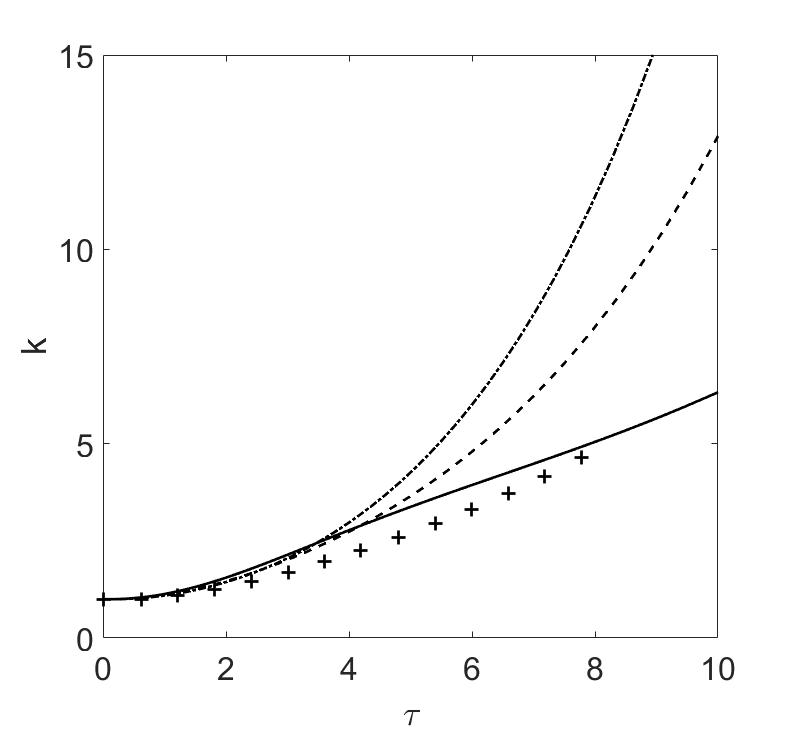}}\\
\caption{Turbulence kinetic evolution for purely sheared flows a) S*=3 b) S*=15 c) S*=27. The predictions of the present model are shown by the solid line, the SSG and the LRR model are shown in dash-dot and dotted lines. The data from the direct numerical simulation of Isaza and Collins \cite{isaza2009} is included for comparison}
\end{figure}

It is documented that the LRR and SSG models may not give satisfactory performance in many elliptic streamline flows. \cite{bns} have simulated homogeneous turbulence subjected to elliptic mean flows:
\begin{equation}
\frac{\partial U_i}{\partial x_j}=
\begin{bmatrix}
  0 & 0 & -\gamma-e \\
  0 & 0 & 0 \\
  \gamma-e & 0 & 0
\end{bmatrix}
\end{equation}
where $e=\sqrt{\frac{1-\beta}{2}}$ and $\gamma=\sqrt{\frac{\beta}{2}}$. For $e>\gamma$ the mean flow has elliptic streamlines of aspect ratio $E=\sqrt{(\gamma+e)(\gamma-e)}$. We use this data from 3 simulations with mean flows having aspect ratios E=3,2 and 1.5. The turbulent velocity field is initially isotropic and the initial $\frac{\eta}{Sk}$ = 0.167. 

\begin{figure}
\centering
\subfloat[$b_{12}, S^*=3$]{\includegraphics[height=5cm]{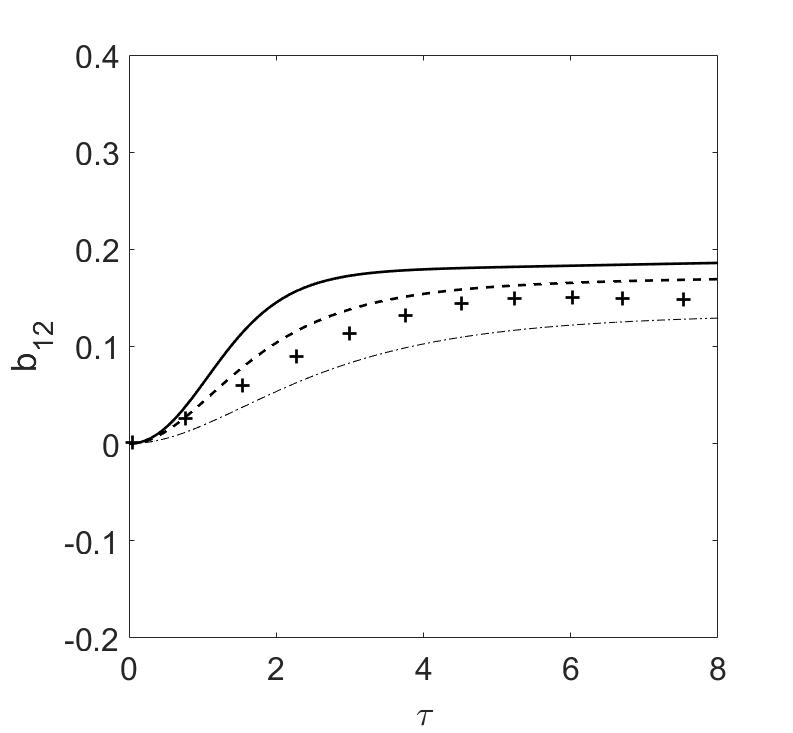}}\\
\subfloat[$b_{12}, S^*=15$]{\includegraphics[height=5cm]{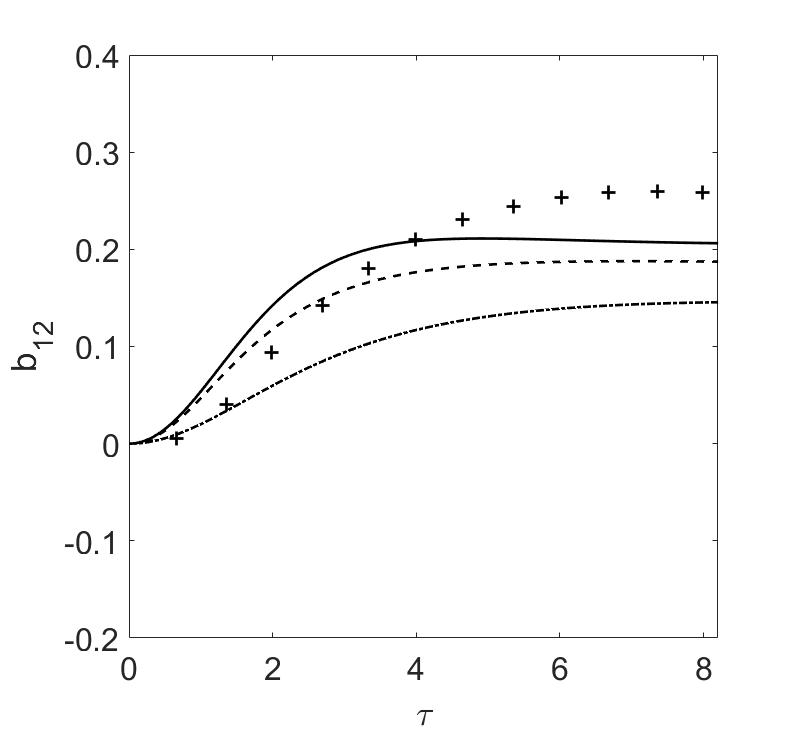}}\\
\subfloat[$b_{12}, S^*=27$]{\includegraphics[height=5cm]{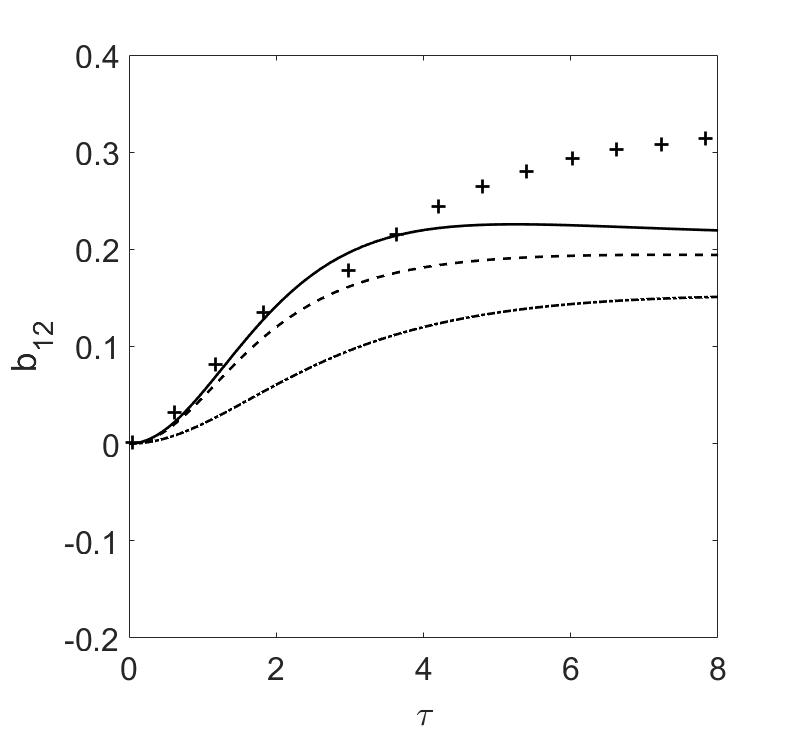}}\\
\caption{Reynolds stress anisotropy $b_{12}$ for purely sheared flows a) S*=3 b) S*=15 c) S*=27. The predictions of the present model are shown by the solid line, the SSG and the LRR model are shown in dash-dot and dotted lines. The data from the direct numerical simulation of Isaza and Collins \cite{isaza2009} is included for comparison}
\end{figure}

\begin{figure}
\centering
\subfloat[$b_{12}, S^*=3$]{\includegraphics[height=5cm]{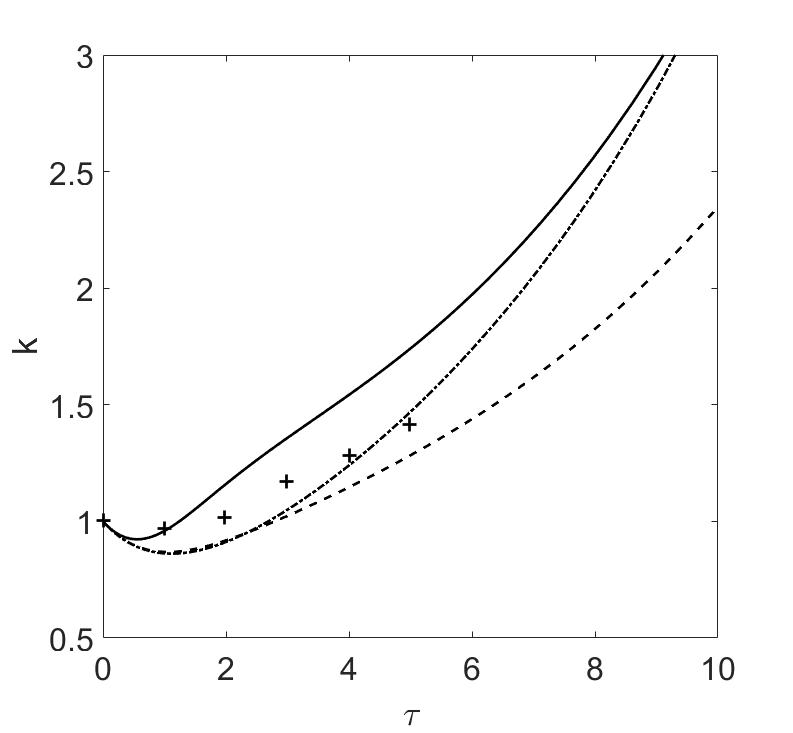}}\\
\caption{Evolution of turbulent kinetic energy for the purely sheared flow. The predictions of the present model are shown by the solid line. SSG and LRR model are shown by the dashed and dash-dot lines. The data from the direct numerical simulation of Bardina et al. \cite{bardina} is included for comparison}
\end{figure}

In Figure 3 and Figure 4 we use the data from the direct numerical simulations of \cite{bns} in elliptic streamline mean flows. Figure 3 represents the time evolution of turbulence kinetic energy for elliptic mean flow with different values of aspect ratio. For case E=1.5 in Figure 3 (a) the LLR and SSG models predict turbulent kinetic energy growth but at a rate much lower than the DNS of \cite{bns}. As the relative strength of mean rotation effect increases, in Figure 3 (b) and (c), the performance of LLR and SSG becomes less satisfactory. For the case E=3 the LLR and SSG models predict turbulent kinetic energy decay but the DNS predicts turbulent kinetic energy growth. For all 3 cases the predictions of the present model are in agreement with the DNS data qualitatively and quantitatively. Unlike LRR and SSG models, the present model predicts growth of turbulent kinetic energy for all three cases of elliptic streamline mean flow. The rate of growth of turbulent kinetic energy predicted by the present model is able to show quantitative agreement with the DNS data also. 

In Figure 4, the evolution of Reynolds stress anisotropy ( $b_{13}$ component) is shown. For all three values of aspect ratio the new model predictions shows improvement agreement with the DNS data of \cite{bns}. Testing across a variety of elliptic streamline flows seems to suggest that the present model is able to show significant improvement in predictions of both the Reynolds stress anisotropy and the turbulent kinetic energy evolution.

In figure 5 and 6, we perform a exhaustive validation for the case of homogeneous sheared mean flow. This flow case is of great importance theoretically and from the point of view of the engineering applicability of the model. We use the data from \cite{isaza2009} where the evolution of the Reynolds stress anisotropies and the turbulent kinetic energy was collected for a range of different shear parameter $S^*$. This is important as it tests the performance of the slow and rapid pressure strain correlation models when used in conjunction with each other. This issue is emphasized in \cite{speziale1992testing} where the authors comment that testing the rapid and slow pressure strain correlation models in isolation can lead to unsound and misleading results. Testing the complete pressure strain correlation model, for a range of  in this manner acts as an exhaustive test of entire pressure strain correlation model as a unit where the rapid and slow models work in conjunction with each other.  We select three specific cases of the shear parameter from \cite{isaza2009}, a) $S^*=3$ b) $S^*=15$ c) $S^*=27$. At $S^*=3$, the nonlinear behavior is dominant in the flow physics and the performance depends more on the accuracy of slow pressure strain model. At $S^*=27$, the linear behavior is dominant and the performance depends more on the accuracy of rapid pressure strain model. Finally, At $S^*=15$, both linear and non-linear physics are of equal importance in the turbulence evolution. This case tests how well the entire pressure strain correlation model works as a unit. The present model predictions matches well with the DNS data for all three values of the shear parameter. There is a significant improvement over the predictions of both the LLR and SSG models.
  
In Figure 7, the present model prediction of the evolution of turbulence kinetic energy is compared with the large eddy simulation data of \cite{bardina} for purely sheared flows. The predictions of the present model are in reasonable agreement with the LES data and show accuracy at par with the models of LLR and SSG.

In testing across these flows we find that the present model is able to show some improvements in accuracy for strain dominated flows like multiple examples of homogeneous shear flow \cite{bardina,isaza2009} and plane strain flow \cite{lee1985}. For rotation dominated flows like those investigated by \cite{bns} the present model shows much improvement over the established models of SSG\cite{ssg} and LRR\cite{lrr}.

\section{Conclusions}

It is accepted in the turbulence modeling community that the pressure strain correlation model is a critical component for the success of the Reynolds Stress Modeling approach. Pressure strain correlation models try to capture the effects of the interaction of fluctuating pressure with the fluctuating rate of strain tensor. Such models try to express the effects of the pressure strain correlation using a tensor basis of local tensors like Reynolds stresses, dissipation and mean velocity gradients. The physics that the pressure strain correlation model tries to capture is non-local due to the non-local nature of pressure. Using a limited set of local tensors to capture this physics leads to limitations in model performance. 
In this investigation we extend the tensor basis used for pressure strain correlation modeling. This set of additional tensors sequentially justified based on physics based arguments. We formulate a model using this extended modeling basis. The present model is tested for a wide variety of turbulent flows and contrasted against the predictions of other popular models  like those by \cite{lrr} and \cite{ssg}. It is shown that the new model provides significant improvement in predictive accuracy.  
We are currently testing this pressure strain correlation model for inhomogeneous turbulent flows where the effects of boundaries and walls are important. This article aims to communicate the promising performance of this model in homogeneous turbulent flows to the turbulence modeling community.

%\begin{acknowledgements}
%If you'd like to thank anyone, place your comments here
%and remove the percent signs.
%\end{acknowledgements}

% BibTeX users please use one of
%\bibliographystyle{spbasic}      % basic style, author-year citations
\bibliographystyle{spmpsci}      % mathematics and physical sciences
\bibliography{asme2e}   % name your BibTeX data base

% Non-BibTeX users please use
%\begin{thebibliography}{}
%
% and use \bibitem to create references. Consult the Instructions
% for authors for reference list style.
%
%\bibitem{RefJ}
% Format for Journal Reference
%Author, Article title, Journal, Volume, page numbers (year)
% Format for books
%\bibitem{RefB}
%Author, Book title, page numbers. Publisher, place (year)
% etc
%\end{thebibliography}

\end{document}